\documentclass[12pt,english]{article}
\usepackage{graphics,cite,amssymb,here}
\usepackage{babel}
\newcommand{\be}{\begin{equation}}
\newcommand{\ee}{\end{equation}}
\newcommand{\bi}[1]{\mbox{\boldmath $#1$}}

\begin{document}
\title{Spin-orbit entanglement in time evolution of radial wave packets 
in hydrogenic systems}
\author{ Marcin Turek \\ {\small\it 
 Institute of Physics, Maria Curie-Sk\l odowska University,
  20-031 Lublin, Poland}
  \and  Piotr Rozmej$\,$\thanks{presenting author}
  \\  {\small\it 
Institute of  Physics,
University of Zielona G\'ora, 65-246 Zielona G\'ora, Poland}
}

\maketitle

\begin{abstract}
 Time evolution of radial wave packets built from the eigenstates of Dirac 
 equation for a hydrogenic systems is considered. Radial wave packets 
 are constructed from the states of different $n$ quantum number and 
 the same lowest 
 angular momentum. In general they exhibit a kind of breathing motion with 
 dispersion and (partial) revivals. Calculations show that for some particular 
 preparations of the wave packet one can observe interesting effects in spin 
 motion, coming from inherent entanglement of spin and orbital degrees of 
 freedom. These effects manifest themselves through some oscillations in the 
 mean values of spin operators and through changes of spatial probability 
 density carried by upper and lower components of the wave function.
It is also shown that the characteristic time scale of predicted effects
(called $T_{\mathrm{ls}}$) is for radial wave packets much smaller 
than in other cases, reaching values comparable to (or even less than) the 
time scale for the wave packet revival.
\end{abstract}

\section{Introduction}\label{Introduction}
For more than fifteen years large efforts 
\cite{parker,averbukh,nauenberg,dacic,peres,bluhmsr,BKP} 
have been made to a detailed
understanding of the quantum dynamics of wave packets in simple systems like H
atoms, hydrogenic atoms as well as in simple molecules.
These theoretical investigations resulted in a good understanding of such subtle
interference effects as collapse, revivals, fractional revivals of wave packets
created in a variety of systems. With development of experimental techniques
allowing to tailor many different desired initial states many signatures of
these phenomena have been observed. A rich survey of early studies 
is given in review papers \cite{alber}.

Radial wave packets (RWP) can be relatively easily excited by short laser pulses
\cite{alber86}.  Their motion, at the beginning, resembles a classical motion.
During later stages of the evolution packets undergo
dispersion and (partial) revivals as well as fractional revivals 
\cite{yeaze89,yeaze90}. The idea of construction of the RWP
in the paper is similar to that presented in articles cited above.
However, we include in our investigation the spin degrees of freedom
which motion for RWP can manifest itself much earlier 
than for other kinds of wave packets. 
The most natural framework for our considerations is using relativistic
wave packets built from the solutions of Dirac equation for hydrogenic systems.
Such approach was used by us in \cite{arvie00tu}, where we investigated circular
wave packets  and in \cite{rozme02}, where 
we considered elliptic wave packets, both in hydrogenic systems. 
Results of \cite{arvie00tu,rozme02,turek02phd} 
show that contributions given by 
small components of wave functions are negligible small. Therefore for time
scales of motion which are relevant one can safely use
an approximation in which the time evolution is calculated with non-relativistic
wave functions and relativistic energies. Such approximation is simpler for
analytical presentation as well as for numerical calculations.
We will use this approximation throughout the paper.

\section{Construction of wave packet}\label{construction}

Assume that just after creation the radial wave packet has the following form 
\be
\label{radini}
  \Psi_r(t=0) = \sum_n w_n \, | n \,  l \, l \rangle 
  \pmatrix{ a \cr  b } \, ,          
\ee
where $|nll\rangle$ are eigenstates of the non-relativistic hydrogenic system
with low angular momentum (usually assumed as $l=1$) and $m=l$.
The weight coefficients 
$w_n=(-1)^n \, c_n= (-1)^n \, (2\pi\sigma^2)^{-1/4}\,
\exp{ \left[ {-(n-n_{av})^2/4\sigma^2} \right] } $ 
are given by a Gauss distribution with mean $n_\mathrm{av}$ and dispersion 
$\sigma$. Distributions of that type describe population of Rydberg states
excited by a short laser pulse. The phase $(-1)^n$ is added to obtain the
initial localization of the wave packet at its external 
turning point.  The spinor 
$\scriptstyle{\pmatrix{ a \cr  b}}$ determines the initial 
direction of the spin.
The WP (\ref{radini}) is an approximation of the full relativistic bispinor
whose initial small components are set to zero.

\section{Time evolution}

After transformation to the basis  $|n,l,j,m_j\rangle$ one obtains 
(with notation $j_>=l+1/2$ and $j_<=l-1/2$)  
\be
|\Psi_{r}\rangle = \sum_{n} \, w_n  
\left\{ \,
  a    \,
     | n, l, j_>,j_>   \rangle + 
  b  \, \left( \,
    \sqrt{\frac{1}{2l+1}} \, | n, l, j_>,  j_< \rangle
  + \sqrt{\frac{2l}{2l+1}} \, | n, l, j_<, j_< \rangle \,
          \right) \, \right\}    \, .
\ee

In the basis  $|n,l,j,m_j\rangle$ time evolution of each state is given 
by an exponential factor $\exp (-iE^{+}_{nl}t/\hbar)$ or
$\exp (-iE^{-}_{nl}t/\hbar)$, where 
$E^{+}_{nl}$ and $E^{-}_{nl}$ are energy eigenvalues for $j_>=l+1/2$
and $j_<=l-1/2$, respectively. Precisely
\be
E^{\pm}_{nl}=m_0c^2 \,\left[1+\frac{(Z  \alpha )^2}{\left[\,
n-j_{\stackrel{>}{\scriptscriptstyle{<}}}-\frac{1}{2} + 
\sqrt{\left( j_{\stackrel{>}{\scriptscriptstyle{<}}}
+ \frac{1}{2}\right)^2-(Z   \alpha )^2}
\,\right]^2}\right]^{-1/2} \, .
\ee
Applying time evolution in that basis and 
transforming back to the   $|n,l,s,m_s\rangle$ basis one obtains the 
wave packet after time $t$ in the form
$
|\Psi_{r}(t)\rangle= \pmatrix{|\Psi_1\rangle \cr |\Psi_2\rangle } \, ,    
$
where the upper $\Psi_1 (t)$ and the lower $\Psi_2 (t)$ component of the spinor 
are given by
\begin{eqnarray}  \label{e3}
\Psi_1 (t) &=& \sum_n \, w_n \,
\left\{
 a   \, \exp{(-iE^{+}_{nl}t/ \hbar )}\, |nll\rangle   \right. \cr
&& \left. \hbox{\hspace{5ex}}+  b \, \frac{\sqrt{2l}}{2l+1} \,
\left( \,
\exp{(-iE^{+}_{nl}t/ \hbar )} -  \exp{(-iE^-_{nl}t/ \hbar )}
\, \right) \, |nll-1 \rangle \, 
\right\} \, ,
\cr 
\Psi_2 (t) &=& \sum_n \, w_n \,
\left\{ \,
  b \, \frac{1}{2l+1} \,
\left( \,
\exp{(-iE^{+}_{nl}t/\hbar )} + 2l \,  \exp{(-iE^-_{nl}t/ \hbar )}
\, \right)  \, |nll \rangle \, 
\right\}  \, .
\end{eqnarray} 
Such wave packet is localized only in the radial coordinate, hence the radial
density probability of the components 
\be
 \rho_1(r) = r^2 \int  d\Omega \, | \Psi_1(r,\theta,\phi)|^2   
 \hbox{ \ \ \ \ and \ \ \ \ }
  \rho_2(r) = r^2 \int  d\Omega \, | \Psi_2(r,\theta,\phi)|^2 \, 
\ee
are convenient quantities illustrating the
wave packet motion. 
The integration over angular coordinates leads to formula
\begin{eqnarray}
 \rho_1(r) &=& r^2 \, \left( \, a^2 \, 
\left| \, \sum_n \, w_n \, R_n(r) \, \exp{(-iE^{+}_{nl}t/ \hbar )}   \, \right|^2 \right. \cr
&& \left.+ \hspace{2.4ex} b^2 \frac{2l}{(2l+1)^2} \,
\left| \, \sum_n \, w_n \, R_n(r) \,
\left( \,\exp{(-iE^{+}_{nl}t/ \hbar )} -  \exp{(-iE^-_{nl}t/ \hbar )} \, \right)  \, \right|^2 \;
\right) \cr
\rho_2(r) &=& r^2 \left(
 b^2 \frac{1}{(2l+1)^2} \,
\left| \, \sum_n \, w_n \, R_n(r) \, 
\left( \, \exp{(-iE^{+}_{nl}t/\hbar )} + 2l \,  \exp{(-iE^-_{nl}t/ \hbar )} \, \right)  \, \right|^2  
  \; \right) \, , \nonumber \\ 
\end{eqnarray}
where $R_n(r)$ denotes  the radial part of the wave function 
$\langle \bi{r}|nlm\rangle$.

\begin{figure}[htb] 
\centering{
\resizebox{0.8\textwidth}{!}{\includegraphics{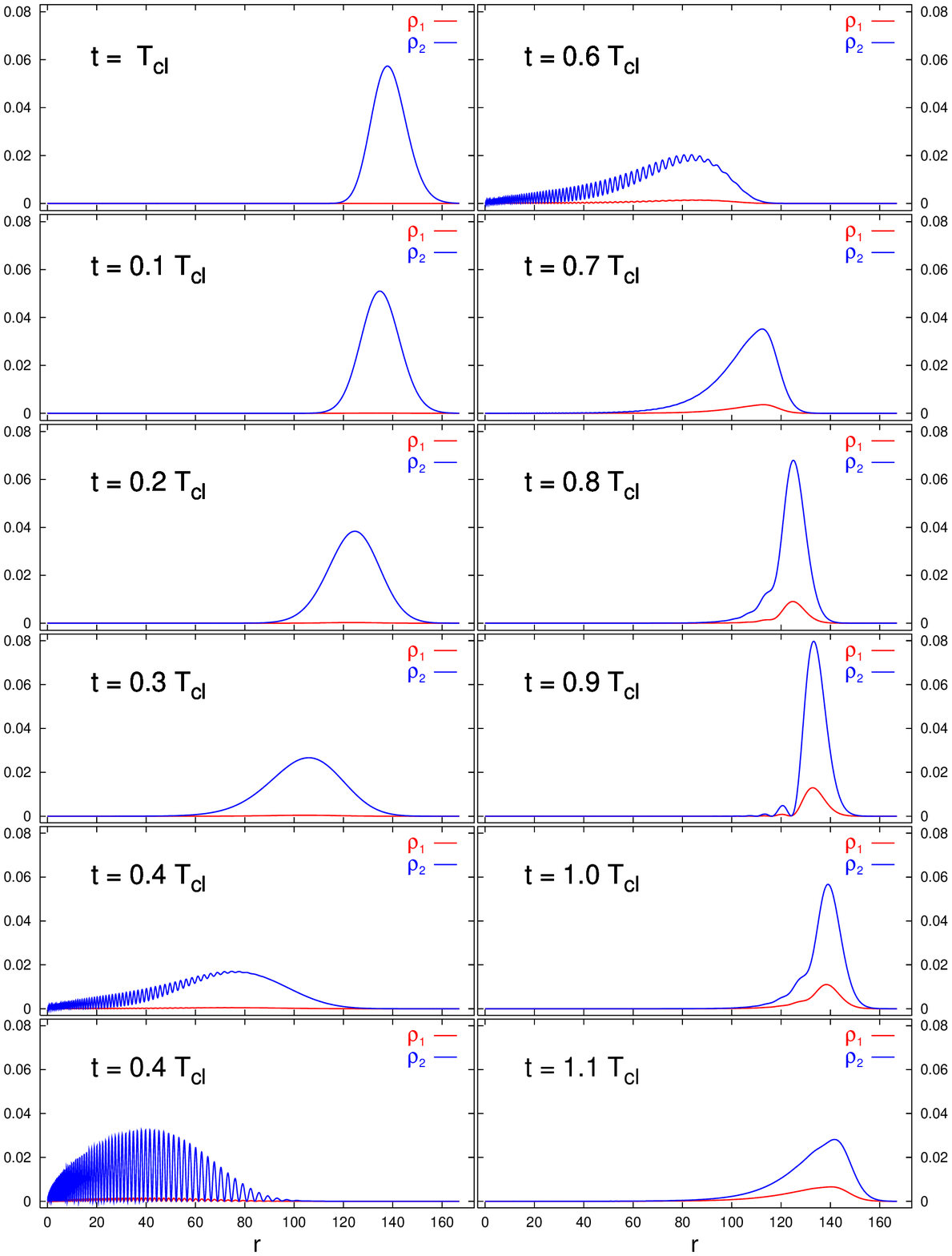}  }}
\caption{  Short time scale evolution of the radial wave packet with 
 $n_{\mathrm{av}}=80,\; a=0,\; b=1$ and $\sigma=2$.  } \label{bclrad}   
\end{figure}

Fig.~\ref{bclrad} illustrates the time evolution of the radial wave packet 
with $n_\mathrm{av}=80$, $a=0$, $b=1$, what corresponds to initial spin 
antiparallel to $Oz$ axis. The wave packet exhibits a kind of breathing motion,
moving towards the center and reassembling itself (approximately) after one
classical period at the external turning point. 

\begin{figure}[htb] 
\centering{ 
\resizebox{0.9\textwidth}{!}{\includegraphics{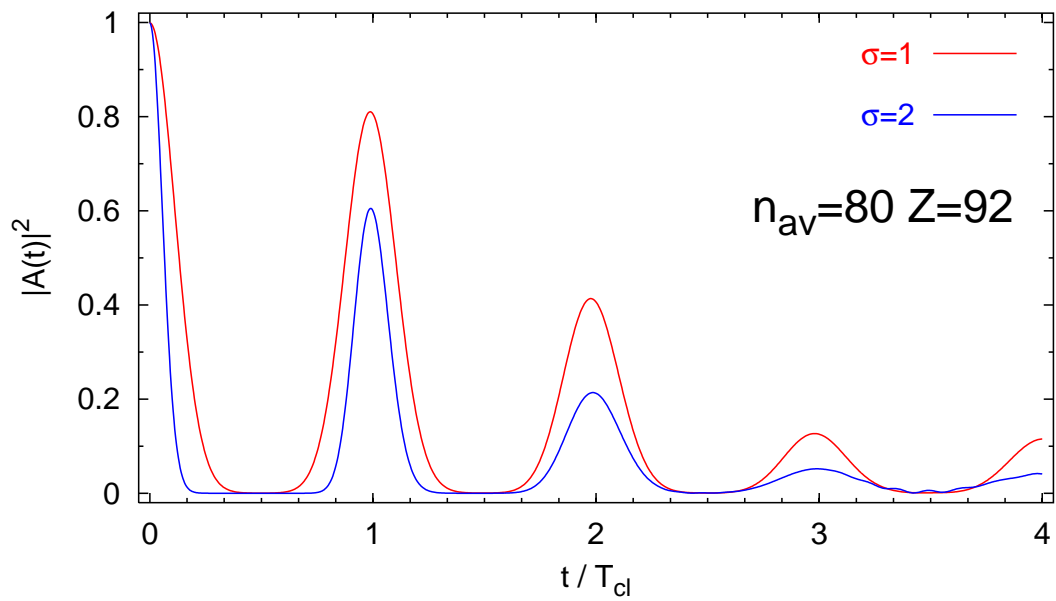}  }}
\caption{ \label{radac}
Autocorrelation function (squared) for radial wave packets with
$n_{av}=80$, $Z=92$,  $\sigma=1$ and 2.  
     }    
\end{figure}

Periodicity of the motion is well seen with the help of the autocorrelation
function $A(t)=\langle \Psi (t)|\Psi (0)\rangle$. For radial wave packet it
reads as
\be 
  A(t) 
= \sum_{nlm}  w_{n}^2 \left\{  \, \left(
  a ^2 \,  + b ^2 \, \frac{1}{2l+1} \right) \, \exp{(-iE_{nl}^+t/ \hbar )}
 +   
     b ^2\, \frac{2l}{2l+1} \exp{(-iE_{nl}^-t/ \hbar )}  \right\} \,  \, .
\ee
The plot of $|A(t)|^2$ for wave packets with $n_{\mathrm{av}}=80,\; Z=92$ 
and two different values of $\sigma$ is presented in Fig.~\ref{radac}. 

Even for short time evolution presented in fig.~\ref{bclrad} 
one can see a transfer
of the probability density from one component of the spinor to the other already
after one classical period. It shows that the time scale of the period of {\em
spin-orbit motion} is for RWP substantially smaller than for
circular or elliptic WP. The period of spin-orbit motion is determined by 
the splitting of energy levels for which $n=n_\mathrm{av}$ 
is maximally populated in WP and
spin projections are opposite
\be
T_{\rm ls} =  \frac{2 \pi \hbar }{|E_{n_{av}l}^+-E_{n_{av}l}^-|} \simeq
\frac{4 \pi  \, n_{av}^3}{Z^4 \alpha^2}=\frac{2l(l+1)} {(Z \alpha)^2}
\, T_{\rm cl} \, .
\ee
The result is obtained using lowest order approximation for relativistic
energies in a hydrogenic system. It is clear that for RWP, whose $l=1$,
$T_{\rm ls}$ can be even smaller than $T_{\rm rev}$, particularly for large $Z$.
 
\begin{figure}[htb]
\centering{ 
\resizebox{0.8\textwidth}{!}{\includegraphics{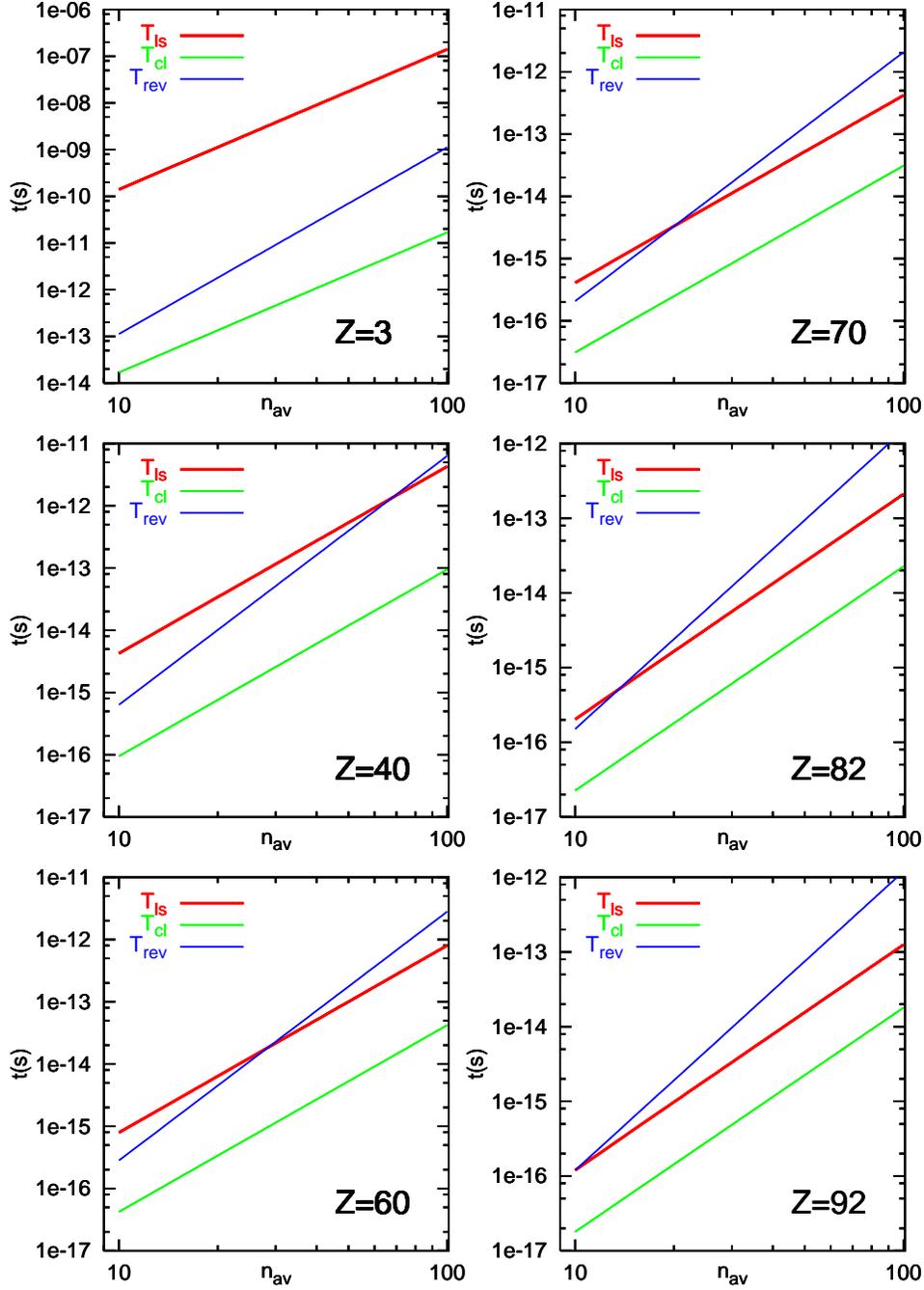}}   }
\caption{ \label{rdts}  Log-log plot of
time scales (in seconds) $ T_{\rm cl}$,  $ T_{\rm rev}$ and $T_{\rm ls}$ of
radial WP as function of $n_{\rm av}$ for different $Z$.
     }
\end{figure}

A hierarchy of time scales is defined as in \cite{arvie00tu}. Writing the
energy as function of single quantum number $n$, for $n=n_\mathrm{av}$ we
define a hierarchy of times 
\be
\frac{1}{k!}\left( \frac{d^k E}{dn^k}\right)_{n=n_\mathrm{av}} =
 \frac{2\pi\hbar}{T_k}, \quad\quad k=1,2,3,\ldots \;.
\ee
For $k=1$ we obtain the classical Kepler time $T_{\rm cl}$, 
for $k=2$ the revival time
$T_{\rm rev}$ and so on. Fig.~\ref{rdts} presents the three time scales 
$T_{\rm cl}$,  $T_{\rm rev}$ and $T_{\rm ls}$ for RWP as functions of 
$n_{\rm av}$ for different $Z$. It is clear that for $Z>60$ $T_{\rm ls}$
becomes comparable to $T_{\rm rev}$ and even shorter for $n_{\rm av}>50$.
Because lifetimes of wave packets with respect to radiative decay are about
two orders of magnitude larger than  $T_{\rm ls}$ 
(according to \cite{chang} $T_{\rm ls}/T_{\rm rad dec}\approx 0.06$)
it is much bigger chance to observe effects of spin-orbit entanglement for RWP
than for any other WP.

\section{Expectation values of spin operators} 

\begin{figure}
\centering{
\resizebox{0.8\textwidth}{!}{\includegraphics{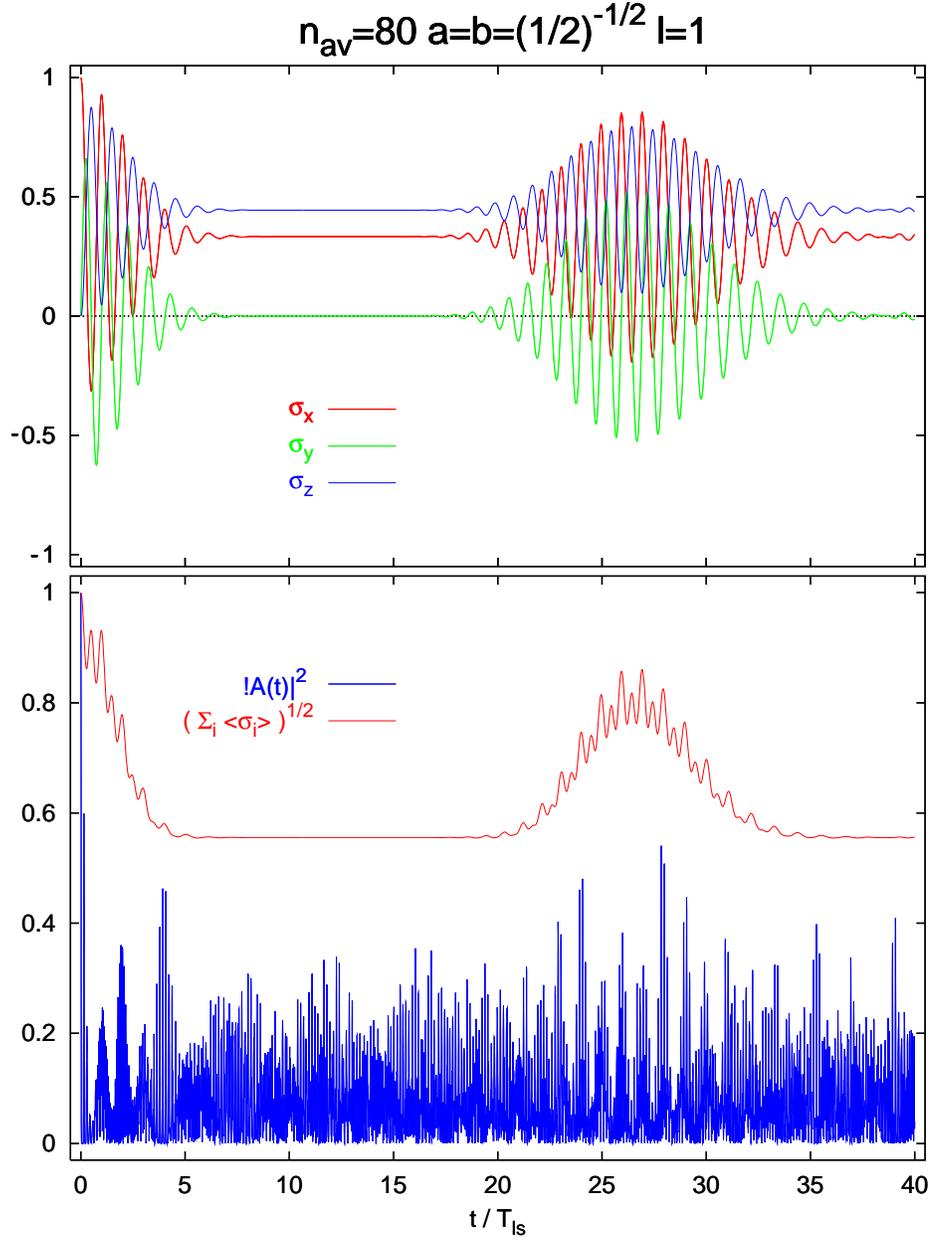}  }  }
\caption{ \label{n80sa}
 Time evolution of spin expectation values for RWP with 
  $n_{av}=80,~l=1$ and $a=b$ (upper part) and the square of the autocorrelation
  function and the 'length' of the spin vector (lower part).        }
\end{figure}

Expectation values of spin operators can be easily obtained 
using equations (\ref{e3}). For RWP they have a simple structure
\begin{eqnarray}
\label{rsx}
\langle \sigma_x \rangle_t &=&  a b \, \sum_n \,|w_n|^2 \,
\left[ \frac{2l}{2l+1}+\frac{4l}{2l+1} \, \cos{ ( \omega_n t ) }
\right] \, ,
\\
\label{rsy}
\langle \sigma_y \rangle_t &=&  a b \, \sum_n \,|w_n|^2 \,
 \frac{4l}{2l+1} \, \sin{ ( \omega_n t ) } \, ,
\\
\label{rsz}
\langle \sigma_z \rangle_t &=& \sum_n  |w_n|^2 \,
\left[ \,
 a^2-  b^2 \, \frac{(2l+1)(2l-1)}{(2l+1)^2} -
 b^2 \, \frac{8l}{(2l+1)^2} \, \cos{( \omega_n t )}
\right] \, , \\
\nonumber
\end{eqnarray}
where $\omega_n = (E_{nl}^+ - E_{nl}^- / \hbar)$. The terms containing 
$\cos {( \omega_n t )}$ and $\sin {( \omega_n t )}$ indicate that at the
beginning one can expect the spin precession followed by the spin collapse 
implied by nonlinear dependence of frequencies  $\omega_n$ on $n$.
This behaviour is clearly seen in the upper part of fig.~\ref{n80sa},
where for $t/T_{\rm ls}\in (0,5)$ spin vector makes several rotations while 
for $t/T_{\rm ls}\in (5,20)$ stays almost constant. At that times the length
of spin vector is reduced to $\bi{\sigma}\approx 0.55$ which means that the
part of spin angular momentum in dynamically transferred into orbital motion.
Later, for $t/T_{\rm ls}\in (20,33)$ spin revives (at half of 
$T_{\rm ls_2}= (2/3)n_{\rm av}T_{\rm ls}$, that is $t\approx 26.7\,T_{\rm ls}$
for $n_{\rm av}=80$).
 The high peak of autocorrelation function and the larger length of spin vector 
 presented in the lower part of fig.~\ref{n80sa} confirm the predicted time of
 the spin revival.

\begin{figure}
\centering{
\resizebox{0.8\textwidth}{!}{\includegraphics{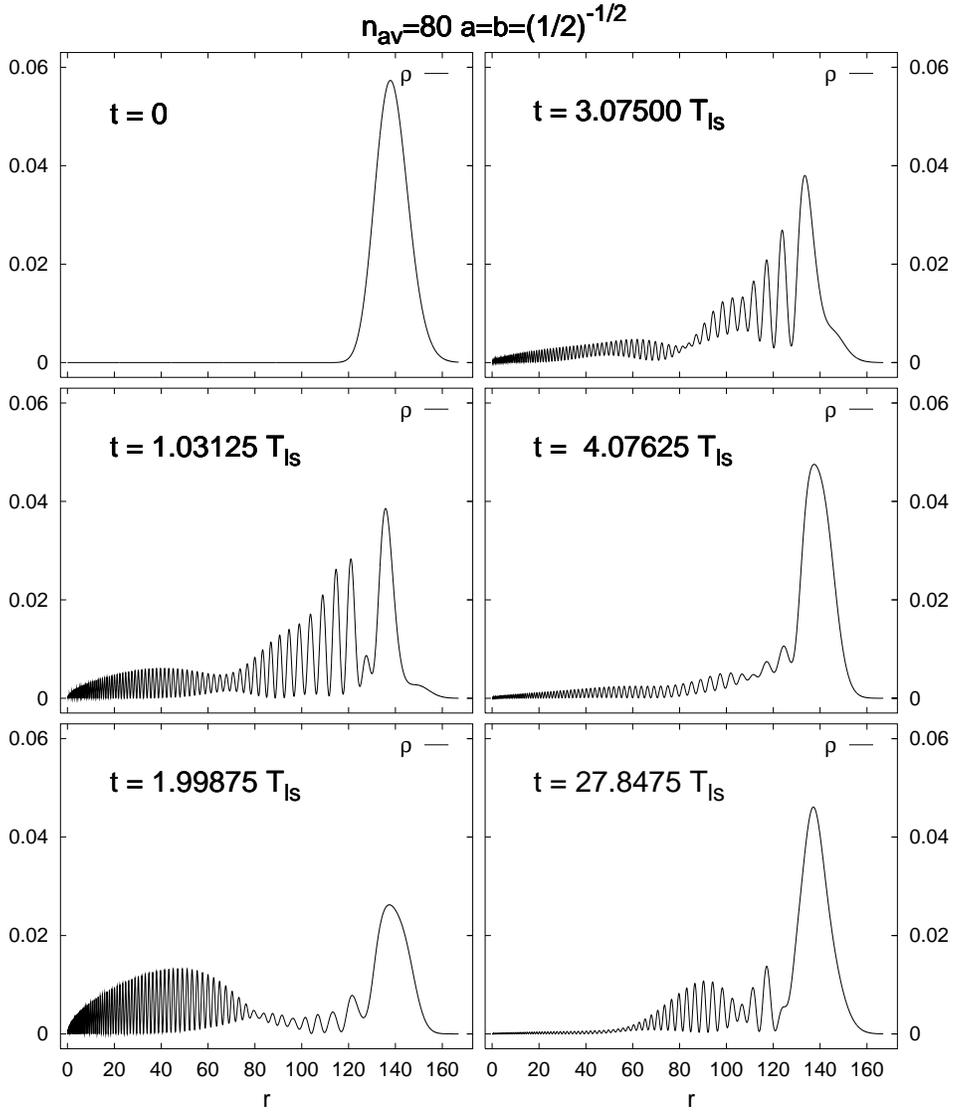}}     }
\caption{ \label{lsr}
 Radial probability density $\rho=\rho_1+\rho_2~$ for RWP with 
 $n_{\rm av}=80,~l=1$ and $a=b$ for several time instants corresponding to big 
 values of autocorrelation function (revivals).  }
\end{figure}

The spin precession is accompanied by the revivals of spatial 
probability density. It is visible
in the autocorrelation function and spin components (fig.~\ref{n80sa})
and in details in fig.~\ref{lsr}  displaying the spatial probability density
for some particular times.

The inherent entanglement of the spatial and spin degrees of freedom manifested
already for short times in fig.~\ref{bclrad} can be illustrated with the help of
{\em quantum carpet} -- space-time plots of the WP evolution \cite{carp}.
Such space-time plot presenting time evolution of 
$\rho_1~\mbox{and}~\rho_2~$ separately is shown in fig.~\ref{cpn80.b.ls1_a}. 
One sees that if the initial WP has only the $\Psi_2$ component there is 
a transfer of probability density to the other component and back.
This transfer is governed precisely by $T_{\rm ls}$ time scale.  

\section{Conclusions}
We have discussed the time evolution of RWP in hydrogenic systems using
a suitable approximation of relativistic approach. The main relativistic effect
is the appearance of the new time scale due to the spin-orbit coupling.
As shown above this time scale can be much smaller for the radial WP than for
previously discussed cases of circular \cite{arvie00tu} or elliptic 
\cite {rozme02} WP. This fact implies that in principle experimental observations of
some spin-orbit effects may become possible with existing techniques.

{\bf Acknowledgement} \\
P.~Rozmej thanks for 
support by the Polish Ministry of Scientific Research and Information
Technology under the (solicited) grant No PBZ-MIN-008/P03/2003.

\begin{figure}
\caption{ \label{cpn80.b.ls1_a}
Time evolution of radial probability densities $\rho_1\,$ and $\rho_2\,$
for RWP with  $n_{\rm av}=80$, $l=1$, $a=0,~b=1$. }
\end{figure}

\end{document}